\documentclass[utf8]{frontiersFPHY} 

\setcitestyle{square} 
\usepackage{url,hyperref,lineno,microtype,subcaption}
\usepackage[onehalfspacing]{setspace}
\usepackage{longtable}

\usepackage{aas_journal_abrv}
\usepackage{hyperref}

\def\keyFont{\fontsize{8}{11}\helveticabold }
\def\firstAuthorLast{Kozarev {et~al.}} 
\def\Authors{Kamen Kozarev\,$^{1,*}$, Mohamed Nedal\,$^{1}$, Rositsa Miteva\,$^{1}$\, Momchil Dechev\,$^{1}$, and Pietro Zucca\,$^{2}$}
\def\Address{$^{1}$Institute of Astronomy and National Astronomical Observatory, Bulgarian Academy of Sciences, Sofia, Bulgaria \\
$^{2}$ASTRON, The Netherlands Institute of Radio Astronomy, Dwingeloo, The Netherlands}
\def\corrAuthor{Kamen Kozarev, 72 Tsarigradsko Chaussee blvd., Sofia 1784, Bulgaria}
\def\corrEmail{kkozarev@astro.bas.bg}

\begin{document}
\onecolumn
\firstpage{1}

\title[Sun to 1 AU SEP Modeling]{A Multi-Event Study of Early-Stage SEP Acceleration by CME-Driven Shocks - Sun to 1 AU} 

\author[\firstAuthorLast ]{\Authors} 
\address{} 
\correspondance{} 

\extraAuth{}

\maketitle
\begin{abstract}
The solar corona below 10 solar radii is an important region for early acceleration and transport of solar energetic particles (SEPs) by coronal mass ejection-driven shock waves. There, these waves propagate into a highly variable dynamic medium with steep gradients and rapidly expanding coronal magnetic fields, which modulates the particle acceleration near the shock/wave surfaces, and the way SEPs spread into the heliosphere. We present a study modeling the acceleration of SEPs in global coronal shock events in the corona, as well as their transport to 1 au, based on telescopic observations coupled with dynamic physical models. As part of the project Solar Particle Radiation Environment Analysis and Forecasting – Acceleration and Scattering Transport (SPREAdFAST), we model the interaction of observed off-limb coronal bright fronts (CBF) with the coronal plasma from synoptic magnetohydrodynamic (MHD) simulations. We then simulate the SEP acceleration in analytical diffusive shock acceleration (DSA) model. The simulated fluxes are used as time-dependent inner boundary conditions for modeling the particle transport to 1 au. Resulting flux time series are compared with 1 au observations for validation. We summarize our findings and present implications for nowcasting SEP acceleration and heliospheric connectivity.

\tiny
 \keyFont{ \section{Keywords:} solar energetic particles, SEP, CME, Shocks, particle acceleration, interplanetary transport} 
\end{abstract}

\section{Introduction}


One of the most common manifestations of solar activity are coronal mass ejections (CMEs). They are usually defined by observations in white light \citep{Vourlidas:2003, Zhang:2006, Bein:2011}, but various aspects of these eruptions are observed in ultraviolet and radio wavelengths \citep{Bastian:2001, Veronig:2010}. In extreme ultraviolet (EUV) light, in particular, the early stages of CMEs may be observed well by telescopes such as the Atmospheric Imaging Assembly \citep[AIA]{Lemen:2012} on board the Solar Dynamics Observatory \citep{Pesnell:2012}. CMEs may drive shock waves in the corona, if their propagation speeds exceed the local speed of information, usually the fast magnetosonic speed. Such shock waves are readily observed in EUV as so-called EUV waves \citep{Thompson:1998}, also known as coronal bright fronts \citep[CBFs]{Long:2011}.

CMEs are the largest contributor to the production of solar energetic particles (SEPs), ions and electrons of energies several orders of magnitude above the thermal coronal plasma \citep{Reames:1999}. Flares are the other important source of SEPs in solar eruptions. SEP production in CMEs occurs mostly in the magnetized shock waves they drive, as well as in plasma compressions caused by the CMEs. Historically, the bulk of the SEP acceleration was thought to occur in interplanetary space, inferred from in situ observations of energetic storm particle (ESP) fluxes during the passage of interplanetary shocks by spacecraft. However, over the last fifteen years more advanced observations and numerical models have confirmed that in their early stages (below 5-10~$R_{Sun}$) CMEs often drive shocks \citep{Ontiveros:2009, Gopalswamy:2011}, and those shocks may accelerate SEPs to energies up to and beyond 100~MeV/n \citep{Battarbee:2013, Kozarev:2013, Schwadron:2014, Kong:2017}. Consequently, recent work has been devoted to characterizing the dynamics of CMEs and the shocks they drive in the corona, using ever more advanced observations in white light, EUV, and radio. 

This has been explored in studies to obtain estimates of the early-stage SEP production in the corona \citep{Kozarev:2013, Schwadron:2015}. In particular, \citet{Kozarev:2019} studied 9 separate western CBF events, and used the diffusive shock acceleration (DSA) model of \citet{Kozarev:2016} to simulate the particle acceleration very early on, while the CMEs were still below 1.5~$R_{Sun}$. They discovered that the SEP production varies among events, and it also changes over the event's duration. In addition, acceleration efficiency depends strongly on the varied coronal environment, through which the shock waves propagate. In this work, we extend the work of \citet{Kozarev:2019}, by modeling the CBF-related shock/compression wave dynamics and particle acceleration out to 10~$R_{Sun}$, coupling the results to a global numerical particle transport model, and comparing the model results to in situ observations. This is a significant improvement to our method, and the first extended validated application of Sun-to-Earth physics-based modeling of SEP acceleration and transport, of which we are aware. We outline the methodology, which is the central part of the Solar Particle Radiation Environment Analysis and Forecasting – Acceleration and Scattering Transport (SPREAdFAST) framework for forecasting SEP events, in Section \ref{S2}. The results of our study are presented in Section \ref{S3}. We discuss them and present a summary in Section \ref{S4}.

\section{Materials and Methods}
\label{S2}

\subsection{Event Selection}
We selected the events on the basis of pre-identified proton events in 17-22 MeV observed by the SOHO/ERNE instrument in the period 2010--2017, obtaining 216 events. Proton events without identified flares and CMEs and without EUV waves (prior the SEP event) were first dropped from the analysis leading to 177 events. Cases without EUV waves or no EUV data, even though they may have identified flare/CME, were also dropped from any further analysis, as the requirement of the SPREAdFAST model is the presence of an EUV wave, leading to 105 events. Several uncertain EUV waves were dropped due to their relevance to a different solar eruption, leading to 99 events. Of those, a final selection of 62 events had measurable near-limb or off-limb CBFs, which can be analyzed with our framework. All events are given in Table \ref{table:1}. Since there are multiple aspects and phenomena related to the events, we provide as reference the date; start, end times, and class of the associated flare; and the source location on the disk in helioprojective Cartesian coordinates. These were obtained from the Heliophysics Events Knowledge Base \footnote{\url{www.lmsal.com/isolsearch}}.

\subsection{The SPREAdFAST Framework}

To model the particle fluxes at 1 au and compare them with observations, we utilize the physics-based global modeling framework Solar Particle Radiation Environment Analysis and Forecasting--Acceleration and Scattering Transport (SPREAdFAST; Kozarev et al. 2021, in preparation). The framework combines detailed EUV observations of CBF events with modeling of the interacting coronal plasma and the resulting SEP production and interplanetary transport. We briefly outline the various components of the SPREAdFAST framework used here.

\subsection{CBF Kinematics and Geometric Modeling}
\label{CBF_kinematics_method}
\begin{figure}[htp]
\begin{center}
\includegraphics[width=16cm]{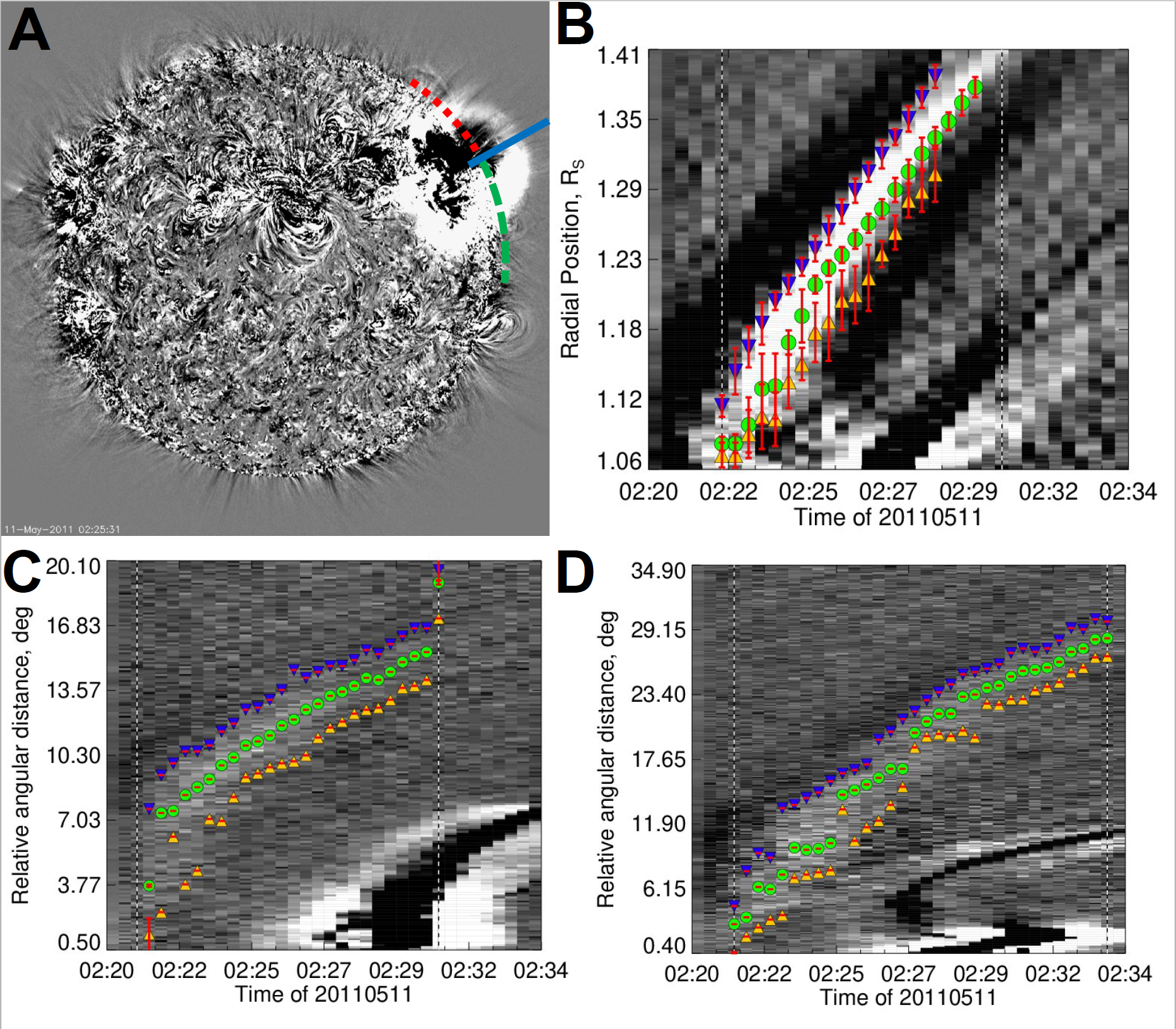}
\end{center}
\caption{Overview of the kinematics measurements for the May 11, 2011 event. \textbf{A.} Global SDO/AIA 193 channel base difference image , showing clearly the CBF in mid-eruption. The radial measurement direction is shown as a blue solid line. The two lateral directions are shown as red/green dotted/dashed arcs. \textbf{B.} The radial kinematic measurements, including the J-map and the CBF average front, peak, and back positions. \textbf{C.} The northward kinematic measurements, including the J-map and the CBF average front, peak, and back positions. \textbf{D.} The southward kinematic measurements and J-map.} \label{fig:110511_kinematics}
\end{figure}

To characterize the kinematics of CBFs, we applied to the AIA observations the methodology of the Coronal Analysis of SHocks and Waves (CASHeW) framework \citep{Kozarev:2017}, updated and implemented in SPREAdFAST. Our framework estimates the CBF kinematics in a similar way to \citet{Long:2021} and \citet{Downs:2021}, by following the leading edge of the front on consecutive images. In addition, we calculate the kinematics of the peak and back edge of the CBFs over time, which allows us to estimate their time-dependent mean intensity and thickness. Figure \ref{fig:110511_kinematics} shows an example of its application to the May 11, 2011 CBF event. In the top left panel (A) is a mid-event AIA 193-channel base difference image, clearly showing the CBF. Overlaid are the radial and lateral lines of measurement. The kinematics are determined using time-height maps (also known as J-maps)\citep{Sheeley:1999}: images created by stacking horizontally in time columns of pixels in a desired direction from a solar image. The shape of a track on these images depends on the direction and speed of the feature measured. The system identifies the radial and lateral wave front positions over time in the J-maps generated with the CASHeW code for each event. In the radial direction (Fig. \ref{fig:110511_kinematics}, panel B), they are measured along the line passing through the solar center and the CBF nose (predominant direction of motion). In the lateral direction (Fig. \ref{fig:110511_kinematics}, panels C and D), parallel to the solar limb, the wave front signatures are measured in two directions away from the radial direction. In some cases, lateral wave signatures are only visible in one direction, or are missing altogether. 

Summary plots of the J-maps with overlaid estimated positions and errors for all events are available at the SPREAdFAST catalog. The lateral measurements in both directions are averaged to form a single lateral kinematics time series for each event. The CBF width and mean intensity in both directions are also recorded. Based on the kinematic measurements, a Savitzky-Golay fit is performed to obtain the kinematics in the AIA FOV, as in \citet{Kozarev:2019}. The smoothed positions were extrapolated to 10 Rs using the analytical model of \citet{Byrne:2013}. 

\begin{figure}[htp]
\begin{center}
\includegraphics[width=12cm]{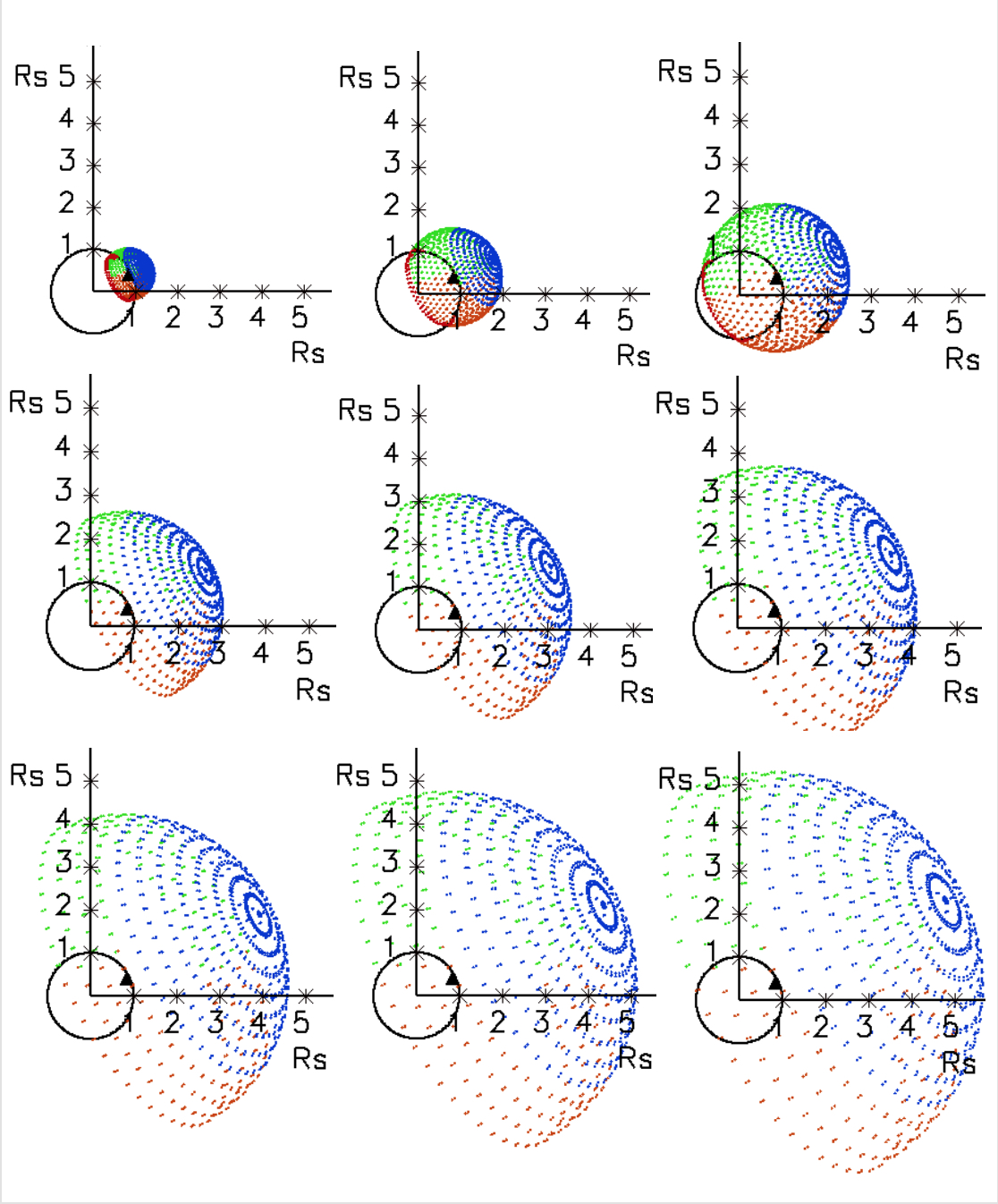}
\end{center}
\caption{An example S2M geometric shock surface model, including the points belonging to the spheroid 'cap' (in blue) and 'zones' (green and red). The footpoints on the Sun are shown in dark red.} \label{fig:s2m_segments_geometry}
\end{figure}

Based on the radial and lateral measured front positions over time, a three-dimensional, time-dependent geometric spheroid model - Synthetic Shock Model (S2M) - was developed for all compressive fronts, consisting of a large number of points ($>$1000), used for the estimation of the dynamic shock upstream coronal parameters. Figure \ref{fig:s2m_segments_geometry} demonstrates 9 time steps of the S2M for an eruption near the northwest limb of the Sun. The spheroid remains centered on the eruption source throughout the event, while its aspect ratio varies based on the radial and lateral CBF position measurements and extrapolations. Once the S2M wraps around the Sun, all points behind the plane passing through the eruption source point, and perpendicular to the radial direction in that point, are removed from consideration. The blue, green, and red points of the S2M in Fig. \ref{fig:s2m_segments_geometry} separate a cap and two side zones, used to investigate the differences in the plasma environment through which the CBF passes in the radial and lateral directions. We have defined the ‘nose’ of the shock model as the collection of all model points on the spheroidal cap subtending a half-angle of Pi/7 from the semi-major axis of the spheroid at each time step. The points on the remaining part of the shock surface is divided into two flanks, or zones, with the dividing plane always parallel to the Sun-Earth line. The plasma parameters at the points on these three surfaces can be examined separately, as demonstrated in Figure \ref{fig:s2m_segments_geometry}.

\subsection{Coronal SEP Acceleration}
\label{section:2.4}
After the plasma parameters along the individual shock-crossing field lines have been established, they are fed in a time-dependent manner into the coronal DSA model \citep{Kozarev:2016, Kozarev:2019} for calculation of the proton acceleration between the low corona and $10~R_{Sun}$. The model was specifically developed to take as input remote solar observations and data-driven model output from the CASHeW framework. The model solves for the coronal charged particle acceleration by large-scale CME-driven shocks. Our model uses time-dependent estimates of shock speed $V_{shock}$, density jump ratio $r$, magnetic field strength $\lvert B\rvert$ and shock angle $\theta_{BN}$, for multiple shock-crossing field lines. The model calculates the minimum shock injection momenta for the particles. It takes as input a particle distribution function, and provides time-dependent distribution function spectra or fluxes, as output. The solution obtained (Eqns. 8-11 in \cite{Kozarev:2016}) gives, for an initial momentum $p_0$, both the first distribution function ($f_1$) and momentum ($p_1$) values, and and an iterative solution for their values ($f_i$ and $p_i$ at subsequent time steps $i$, separated by the observational cadence $\delta t$ of the instrument (SDO/AIA, in this case). The model is run for each individual shock-crossing field line, based on observed and calculated parameters at a single shock-crossing point along it. Flux spectra at each time step are then computed. It has been validated and used for the analysis of a number of SEP events.

We use as input to the model observations-based suprathermal proton spectra derived from 1 au fluxes with the SOHO/ERNE instrument \citep{Torsti:1995}, observed during the 24 hours of quiet time preceding each SEP event. We fit power laws to each suprathermal spectrum in the energy range $0.056-3.0~MeV$, and scale them to $1.05~R_{Sun}$, assuming a simple inverse square dependence on radial distance (implying flux conservation). We currently do not consider adiabatic cooling or other particle transport effects; these are indeed important, but a full study of these effects on suprathermals must be performed in a future study, in order to determine a general trend to use for forecasting. This produces time-independent power law input spectra for the DSA model. The suprathermal spectrum calculated for $1.05~R_{Sun}$ is injected at all shock positions and distances - in the current implementation it is not modified to account for the changing shock locations.

\subsection{Interplanetary SEP Transport}
The final step of the modeling chain is the transport of the accelerated SEPs to 1 au, and subsequent comparison with particle observations with the ERNE instrument. This is achieved by taking the resulting averaged fluxes for the entire event (here an example with the event on May 11, 2011 is shown) as input to the modified version of the EPREM model \citep{Schwadron:2010}, and transporting them through a Parker-type static interplanetary medium. The particle injection from the DSA model into ERPEM is continued over the duration of the coronal shock event. The model includes the effects of pitch-angle scattering, adiabatic focusing and cooling, convection and streaming, and stochastic acceleration. The solver requires inner boundary conditions, with no initial conditions imposed. It features a dynamical simulation grid, in which the computational nodes are carried away from the Sun (frozen-in) with the solar wind - thus the connected grid nodes (streamlines) naturally assume the shape of a three-dimensional interplanetary magnetic field, along which energetic particles propagate. By default, EPREM incorporates an interplanetary magnetic field model with a radial field component falling off as the inverse square of radial distance, azimuthal component falling off as the inverse of radial distance, and a constant latitudinal component - the so-called Parker spiral model. The spatial grid is housed in a data structure based on nested cubes, whose surfaces are regularly subdivided into square arrays of square cells. These cells represent the structure of the grid, within which computational nodes propagate. The inner boundary surface rotates with the solar rotation rate, and is expelled outwards at the solar wind speed. At each time step, a new shell of cells is created at the inner boundary of the grid, and starts its propagation outwards. The inner boundary for the EPREM simulation is at 1.05$R_\odot$. The outer boundary varies for the individual field lines due to the varying dynamic conditions, but it always exceeds 1 AU. The model has been extensively validated and used for SEP studies \citep{Kozarev:2010,Schwadron:2014}.

For the EPREM model runs we performed on the 62 events here, we used the same input parameters, namely: mean free path $\lambda_{0}=0.1$~au at 1~au and 1~GV magnetic rigidity; constant solar wind $V_{sw}=500$~km/s; proton number density at 1~au $n=5.0$~cm$^{-3}$; magnetic field magnitude at 1 au of $\lvert B \rvert=5.0\times10^{-5}$~G. The mean free path is additionally scaled with proton rigidity and radial distance from the Sun to reflect the magnetic turbulence spectrum and its radial dependence \citep{Zank:1998, Li:2003,Sokolov:2004,Verkhoglyadova:2009}:
\begin{equation}
\lambda_{||}= \lambda_{0} \left(\frac{pc}{1 GeV}\right)^{1/3} \left(\frac{R}{1 AU}\right)^{2/3},
\label{eqmfp_vs_r}
\end{equation}
giving the parallel mean free path for the simulation.

A 20-point energy grid was used with equal log-spacing between 1.0~MeV and 200.0~MeV, as well as a 4-point pitch-angle grid. A simulation time-step of 0.5~au/c (approximately 4 minutes) was used with 30 sub-steps, allowing for the proper calculation of the SEP propagation among nodes. All effects of diffusive transport were included in these baseline simulations (adiabatic cooling/heating, adiabatic focusing, pitch-angle scattering, convection with the solar wind, streaming). The effects of perpendicular diffusion and particle drifts will be included in subsequent work. The simulations were stopped at 9.6 hours from the onset of the event at the Sun for all events, since the goal is to model their initial stages.

\section{Results}
\label{S3}
We have modeled 62 events with the SPREAdFAST model framework. To our knowledge, this physics-based Sun-to-Earth study is unprecedented in terms of the sheer number of events modeled from telescopic observations to in situ fluxes directly comparable to observations. All these events have related SEP flux increases at 1~au. Summary plots for each of the modeled events are available at the catalog web page of SPREAdFAST \footnote{\url{http://spreadfast.astro.bas.bg/catalog}}. Below we showcase the results for a single simulated event, and give summarized results for all 62 events.

\subsection{Plasma Parameters Along Individual Shock-Crossing Magnetic Field Lines}

\begin{figure}[h!]
\begin{center}
\includegraphics[width=14cm]{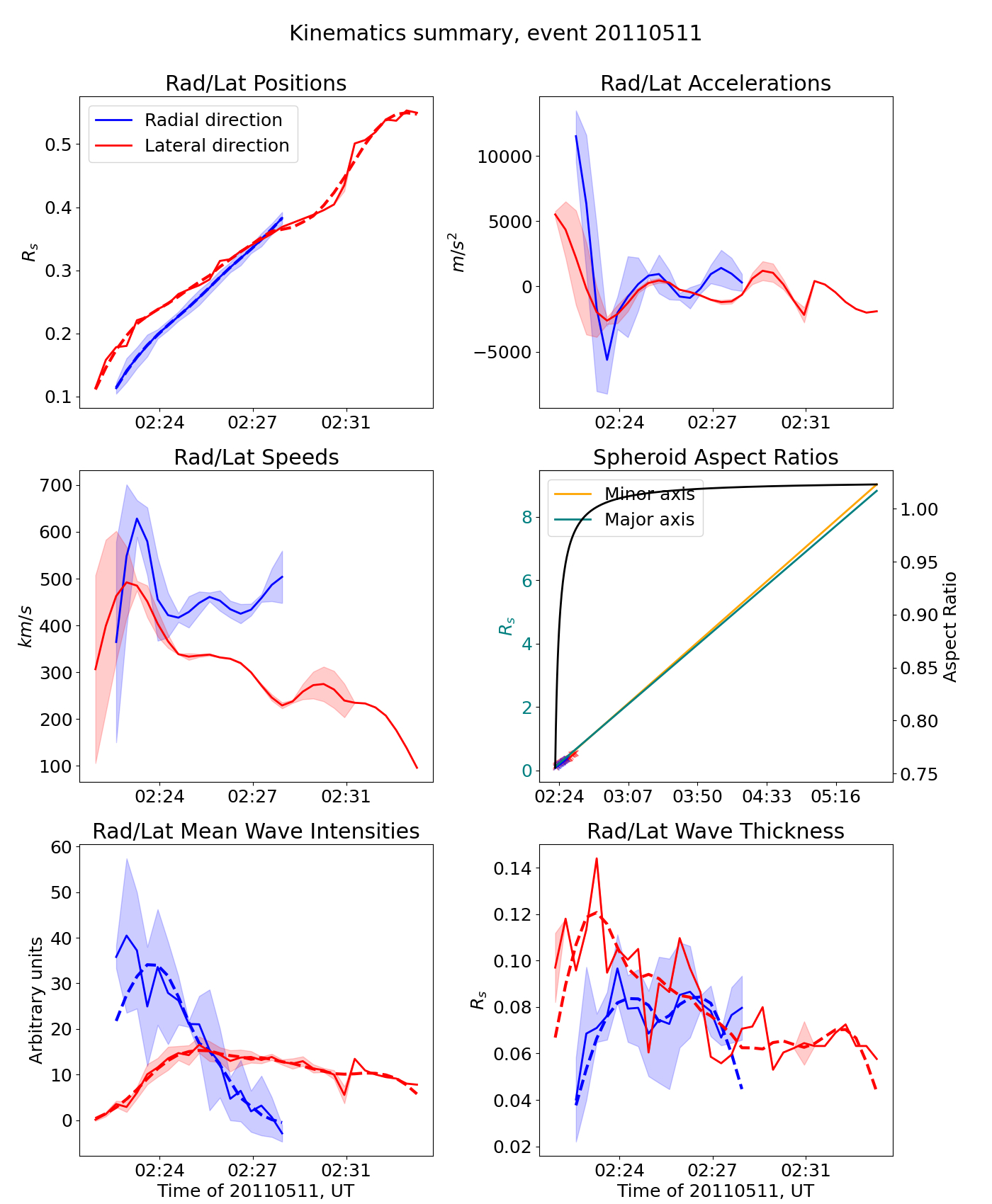}
\end{center}
\caption{Radial and lateral kinematics of the May 11, 2011 SPREAdFAST event. Shown are the radial (in blue) and lateral (in red, mean of the northward and southward directions) time-dependent front positions, as well as calculated speeds and accelerations. Also shown are the mean intensities (calculated as the mean value of the intensity in the J-maps between the SPREAdFAST-determined front and back of the CBFs) at each time step. The CBF thickness in each measurement and direction is also calculated from the automatically determined front and back of the CBF.}
\label{fig:s2m_dynamics_110511}
\end{figure}

For all events in the sample, the kinematics in the radial and lateral directions were obtained using the procedure described in Sect. \ref{CBF_kinematics_method}. As an example, Figure \ref{fig:s2m_dynamics_110511} shows a summary plot of the kinematics characterization for a single event, which occurred on 11 May, 2011 (denoted as 110501\_01). The top left panel shows the radial (blue) and lateral (red) positions above the solar surface, while the other panels show the derived speeds, accelerations, mean wave intensities, and wave thickness (both calculated by using the SPREAdFAST-determined CBF front and back positions in the J-maps). The middle-row right plot shows the extrapolations of the major and minor axes in orange and blue, while the aspect ratio evolution is shown in black. Uncertainties are shown as light blue and light red shadings.

Based on the extrapolated radial and lateral kinematics results obtained for each event, and the inferred major and minor axes of the spheroid representing the shock wave, a three-dimensional geometric model is created, which describes the shock surface at 24-second intervals from the onset of the CBF to the time when its nose (leading direction's front position) reaches 10~$R_{Sun}$. This model shock surface is then propagated through the solar corona up to 10~$R_{Sun}$, represented by the domain of the results of a Magnetohydrodynamic Algorithm outside a Sphere \citep[MAS]{Mikic:1999} synoptic coronal magnetohydrodynamic (MHD) model for the Carrington rotation of each event. The MAS coronal model results are freely available on Predictive Science Inc's MHDweb service\footnote{\url{http://www.predsci.com/mhdweb/}}. The shock surface thus samples at discrete points the relevant parameters for coronal shock acceleration of SEPs, determined by the magnetic field lines crossing it consecutively at each point. 

We sample on the order of 1000 shock-crossing field lines for each modeled event, obtaining time-dependent estimates for the shock-field angle $\theta_{BN}$, shock upstream magnetic field amplitude $\lvert B \rvert$, shock speed $V_{shock}$, plasma density $n$, and Alfven speed $V_A$. The shock density jump, $r$, is determined by estimating for the emission measure and the average density at the shock crossing points at each time step, and dividing by a pre-event average density model. The emission measure and the average densities within the field of view of the AIA instrument are estimated by applying a differential emission measure (DEM) model \citep{Cheung:2015} to each group of pixels surrounding the projected shock-field crossing points for every timestep.

\begin{figure}[h!]
\begin{center}
\includegraphics[width=14cm]{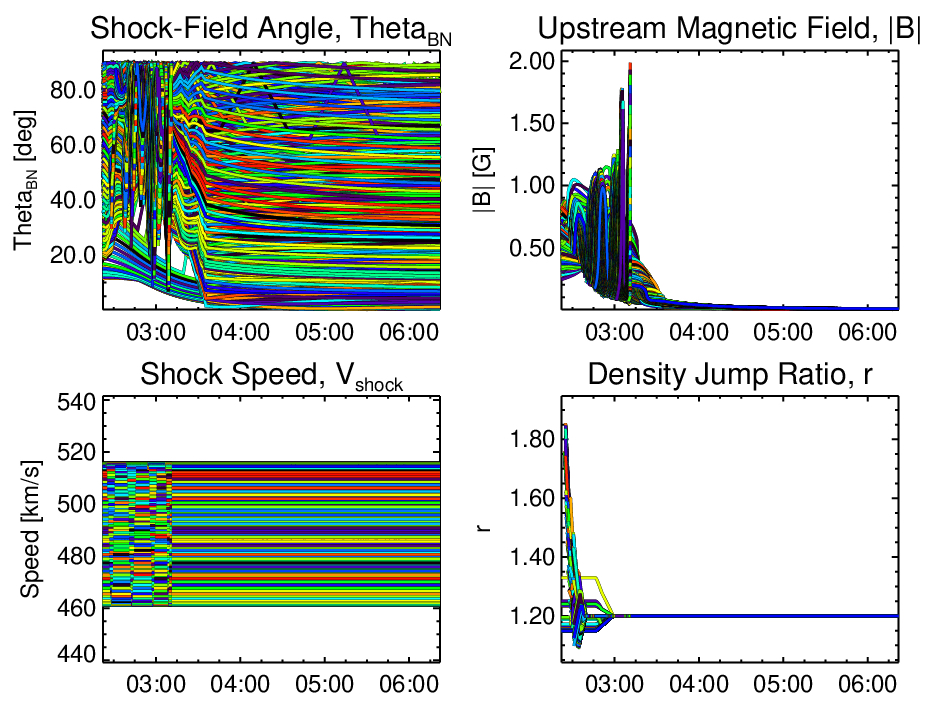}
\end{center}
\caption{Time series of the main parameters used for diffusive shock acceleration modeling, estimated at all shock-crossing field lines for the May 11, 2011 event. Note that the density jump ratio, $r$, is set to 1.2 (representing a weak shock) beyond the field of view of AIA. The different colors are only used to discern between the separate shock-crossing points, and do not carry any additional meaning in this Figure.}
\label{fig:110511_01_shock_parameters}
\end{figure}

\begin{figure}[h!]
\begin{center}
\includegraphics[width=14cm]{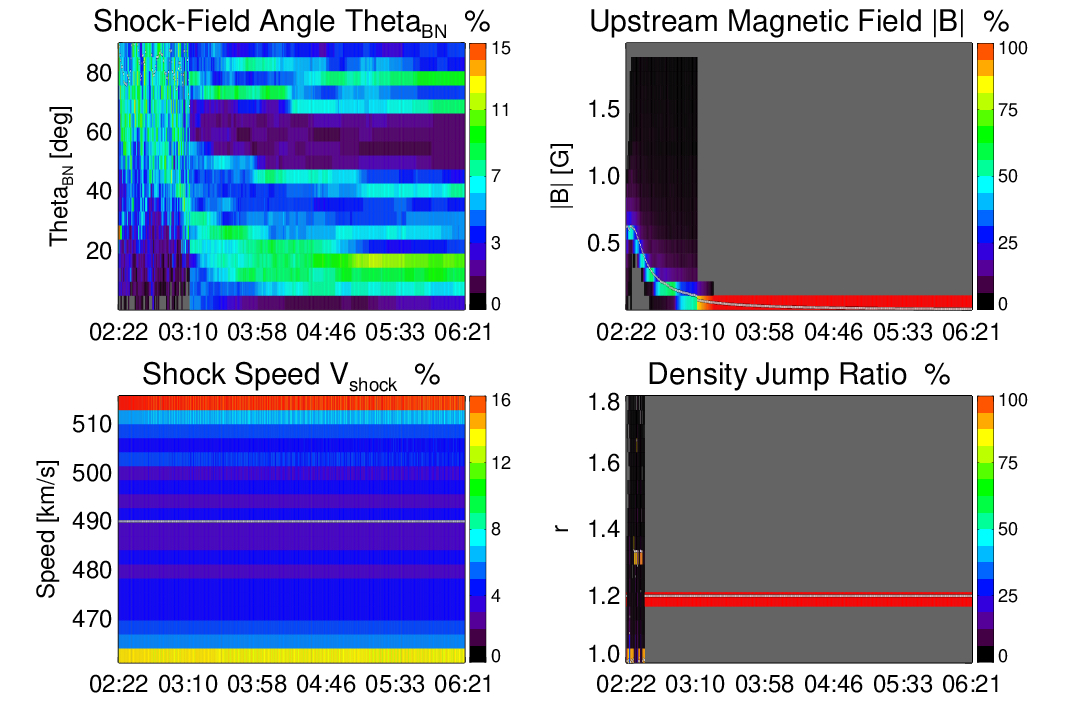}
\end{center}
\caption{A dynamic spectrum of the main parameters used for diffusive shock acceleration modeling, estimated at all shock-crossing field lines for the May 11, 2011 event. This is a summary of the result presented in Fig. \ref{fig:110511_01_shock_parameters}.}
\label{fig:110511_01_shock_parameters_histogram}
\end{figure}

Figure \ref{fig:110511_01_shock_parameters} shows parameter evolution along the individual shock-crossing field lines, for the May 11, 2011 event, between the solar surface and 10~$R_{Sun}$. The panels show, in different colors, the shock-field angle $\theta_{BN}$, shock upstream magnetic field amplitude $\lvert B \rvert$, shock speed $V_{shock}$, and the shock density jump ratio, $r$. Similarly to other authors, \citep{Frassati:2019, Frassati:2020, Long:2021}, we employ DEM analysis to estimate the shock strength from AIA observations. We find that, while we use a different DEM model than these authors \citep{Cheung:2015}, the general results are similar - we find weak coronal shocks. The advantage of our method is that we estimate the density change at all points of the shock model, not only one or several regions. This method has previously been employed in \citet{Kozarev:2017}. We find that the density jump based on the DEM modeling within the AIA field of view is generally quite small, predominantly below 1.2, which is consistent with many previous studies \citep{Kozarev:2015, Vanninathan:2015}. Consequently, beyond the AIA FoV, where we do not have observational information, we set the value to 1.2.

A better cumulative view of the evolution of these parameters for this event is shown in Figure \ref{fig:110511_01_shock_parameters_histogram}. It shows the same four parameters from Fig. \ref{fig:110511_01_shock_parameters} as dynamic spectra over all shock-crossing field lines, with the colors scaled to the minimum/maximum percentages of shock-crossing points in the corresponding parameter value bins for each timestep. This representation shows better the instantaneous distribution of the DSA-relevant plasma parameters at the shock front. For example, the $\theta_{BN}$ and $\lvert B\rvert$ panels shows a consistent decline in the values as a function of time (and radial distance).

\begin{figure}[h!]
\begin{center}
\includegraphics[width=12cm]{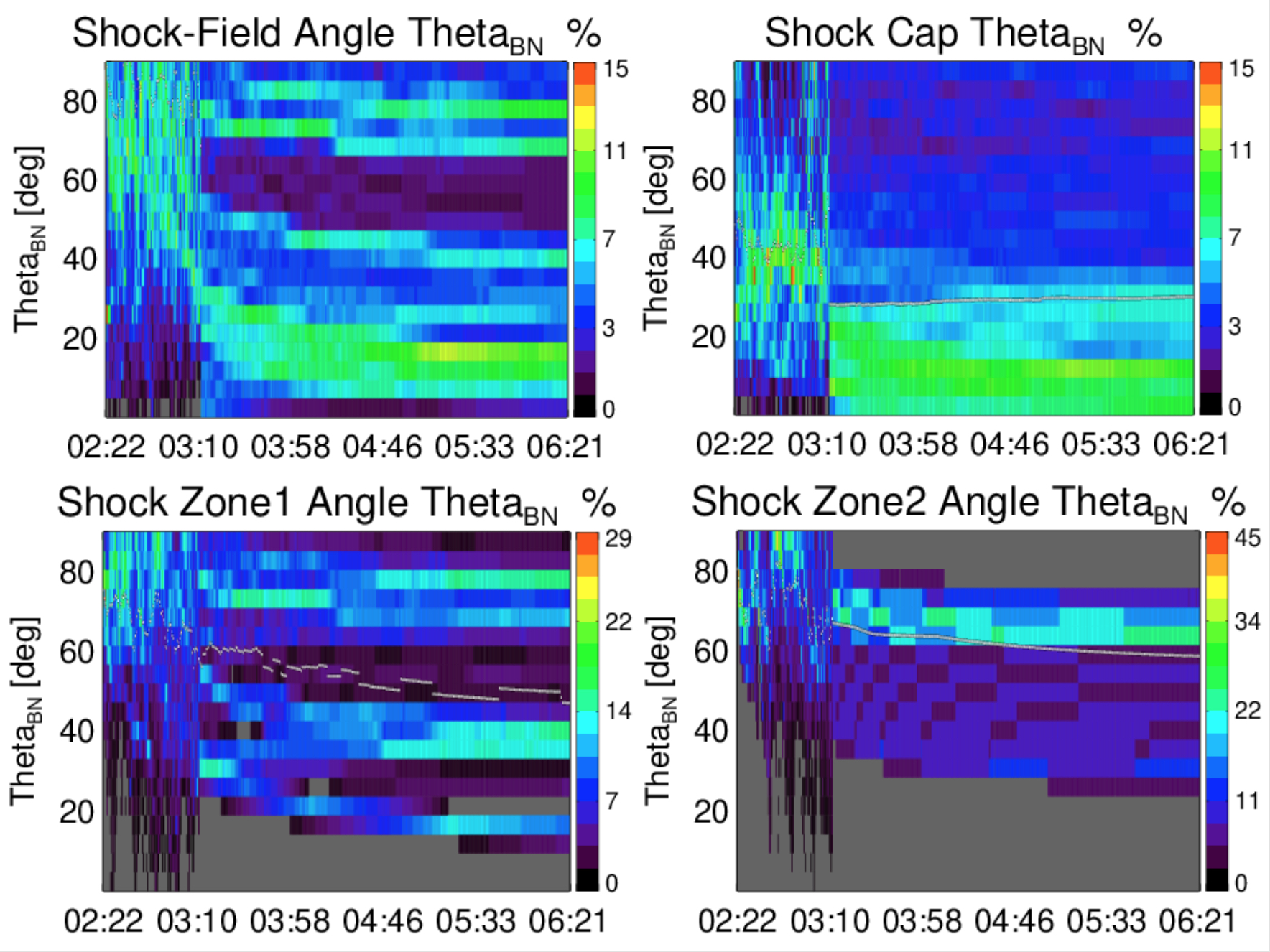}
\end{center}
\caption{Dynamic spectra showing the distributions of the $\theta_{BN}$ shock-field angle as a function of time for the May 11, 2011 event. The top left panel shows the distribution over time for the entire S2M surface. The top right and bottom plots show the distributions for the S2M points of the cap, zone 1, and zone 2, shown in Fig. \ref{fig:s2m_segments_geometry} with blue, green, and red, respectively. The white dot symbols show the mean at each time step.}
\label{fig:thetabn_comparison_110511_01}
\end{figure}

Perhaps the most important parameter for diffusive shock acceleration is the angle $\theta_{BN}$. A clearer picture of its evolution in the model corona for this particular event emerges when the values at each timestep are aggregated into histograms, and shown as a dynamic spectrum, as in Figure \ref{fig:thetabn_comparison_110511_01}. The top left plot shows the distribution as a function of time for the entire spheroid surface. The color table represents the percentage of the total occurrences in each bin at each timestep. The overall evolution for this event is from high to low $\theta_{BN}$ angle, with a significant decrease occurring in the first 50 minutes of the event. The detailed dynamic spectra of the shock plasma parameters, as well as all results shown for individual events below, are available on the SPREAdFAST catalog web page.

A more nuanced picture may be observed if the full surface of the S2M spheroid is logically divided into points belonging to a cap (nose) and two zones (flanks) to the north and south - and the $\theta_{BN}$ time-dependent distribution is extracted separately for each region, as in Fig. \ref{fig:s2m_segments_geometry}. It shows the distribution of points in separate snapshots of the geometric model, with the cap in blue color, zone 1 in green, and zone 2 in red. These correspond to the top right and bottom plots in Figure  \ref{fig:thetabn_comparison_110511_01}. The lowest values of $\theta_{BN}$ are seen in the cap/nose region, while in zone 2 the high values dominate throughout the event, with means consistently above $60^{\circ}$.

We note that it is currently not possible for our model framework to account for the changes occurring in the corona between two closely separated in time events (i.e. two events originating from the same active region in two consecutive days), because it relies on a synoptic time-independent MHD model of the corona. The difference between such historical events with respect to modeling is mainly in the kinematics and shape of the compressive front, which however is expressed in differences in the shock-crossed field lines, shock-field angle $\theta_{BN}$, etc. This naturally has an effect on the modeled fluxes. The observed fluxes from such events also differ. In future work we will strive to incorporate daily updated coronal plasma models, in order to relax the assumption of no short-term change in the coronal conditions.

\subsection{Diffusive Shock Acceleration of Protons}
\begin{figure}[h!]
\begin{center}
\includegraphics[width=14cm]{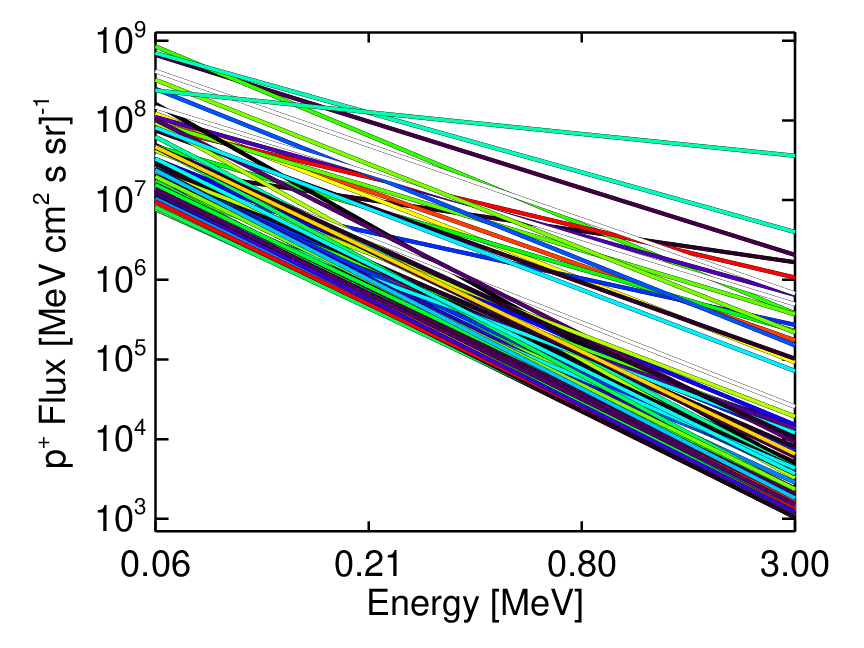}
\end{center}
\caption{The event input spectra for the DSA model, derived from 1 au quiet-time suprathermal observations.}
\label{fig_suprathermal_input}
\end{figure}

The suprathermal flux spectra derived from observations along the method described in Section \ref{section:2.4} for all events are plotted in Figure \ref{fig_suprathermal_input}. It shows a variety of power law indices, dominated by softest spectra. The corresponding time-constant flux spectrum for each event is extended to the range 0.056-200~MeV, and is fed as initial condition into the DSA model of \citet{Kozarev:2016} throughout its spatial domain (1.05-10~$R_{Sun}$). The model's energy grid contains 30 energy steps in this range. For this study, we set a constant upstream scattering mean free path in the corona mfp=0.0055~$R_{Sun}$. The DSA model produces time-dependent distribution function histories for all shock-crossing field lines, which are then aggregated to calculate the average flux time series for each event as a function of energy.

\begin{figure}[h!]
\begin{center}
\includegraphics[width=12cm]{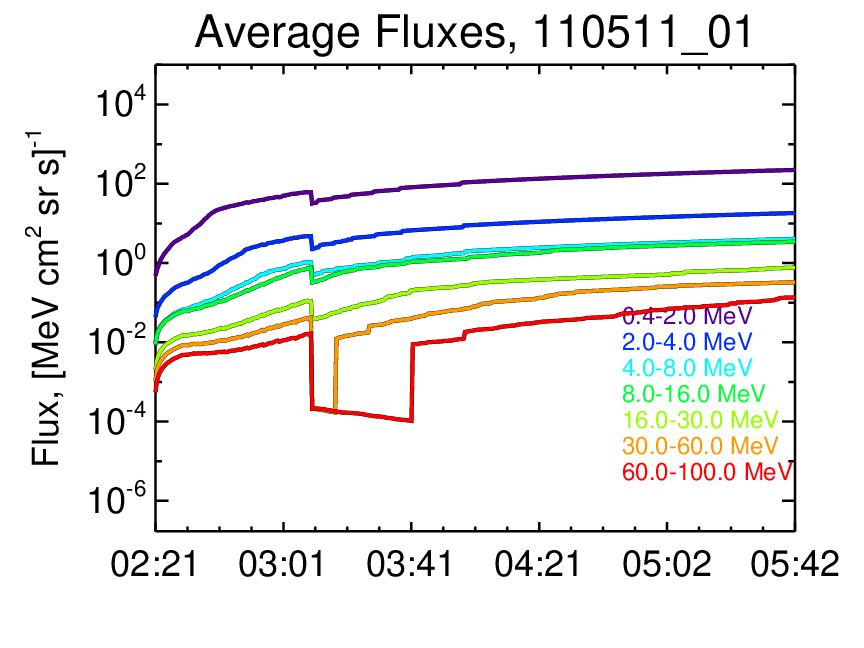}
\end{center}
\caption{Differential proton fluxes simulated with the DSA model for the  event, aggregated in 7 energy channels between 0.4 and 100~MeV. The sharp drop in the fluxes is due to the sudden change of the geometric model between a spheroid and a partial spheroid around 03:10.}
\label{fig_coronal_dsa_fluxes_110511}
\end{figure}

The coronal DSA modeling results for the  event are shown on Figure \ref{fig_coronal_dsa_fluxes_110511}. The plot shows flux time series at seven energy ranges between 0.4 and 100~MeV, obtained by summing the resulting fluxes over all shock-crossing points. The simulation covered over three hours of the event, until the shock reached 10~$R_{Sun}$. The time series show clearly the time-varying (mostly increasing) fluxes, with the first 90 minutes being most dynamic. The sudden drop in flux around 3:10 is due to the shift from the fully-connected spheroid to the reduced surface shock model (i.e. from step 5 to 6 in Fig. \ref{fig:s2m_segments_geometry}), and the consequent reduction in the number of shock-crossing field lines.

\begin{figure}[h!]
\begin{center}
\includegraphics[width=12cm]{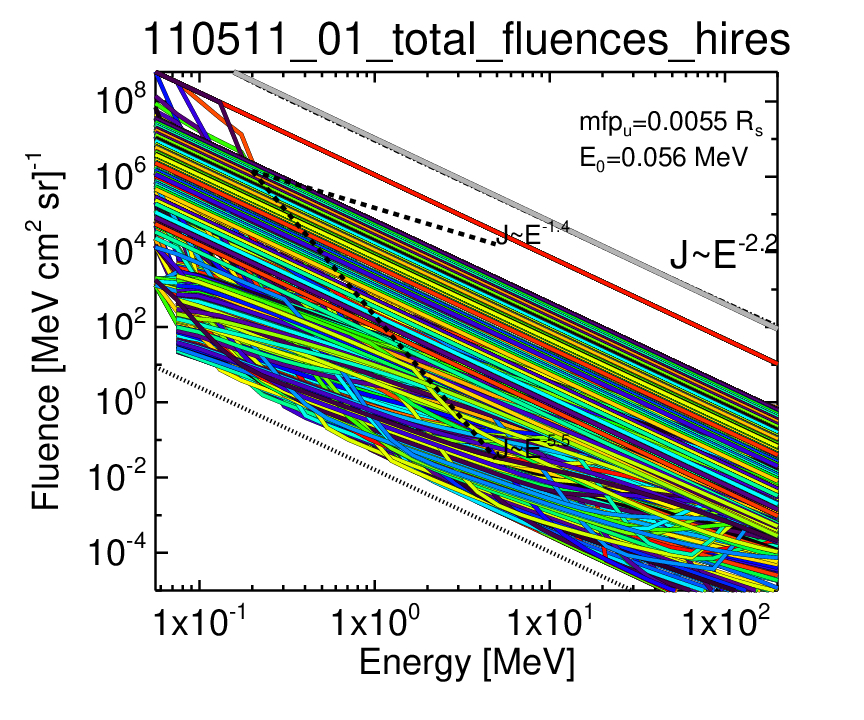}
\end{center}
\caption{Resulting fluence spectra for all shock-crossing field lines modeled with the SPREAdFAST DSA model for the May 11, 2011 event. The sum total of all fluences is also plotted with a light gray color, and the power law index to its fit is shown (in this case, $\delta=-2.2$.) We also show the input spectrum fluence with a dotted line below the colored output fluence lines. The plot also features reference power law slopes corresponding to shock density jumps $r=1.2$ and $r=4$.}
\label{fig_coronal_dsa_fluences_110511}
\end{figure}

The integrated spectra of all individual modeled shock-crossing field lines for event are shown in Figure \ref{fig_coronal_dsa_fluences_110511}. The fluences were obtained by integrating over the time period the individual line fluxes. The energy extent is 0.056-200~MeV, and the overall combined slope is -2.2. The sum of all fluences is also plotted in light gray color. This figure illustrates the variety of accelerated proton fluences obtained due to the different coronal conditions in the simulation. In the interplanetary simulation, the average of these fluences is used. The plots show significant acceleration to even the highest energies for most lines. Realistic fluences in acceleration under time-changing shock plasma parameters do not have to be power laws, that is why some fluence lines have more complex shapes. We note that the S2M model provides upwards of 10 000 shock-crossing field lines; for manageability reasons the framework randomly selects a subset of $\sim$1000 lines for input to the DSA model.

\subsection{Interplanetary SEP Transport Results and Comparisons to In Situ Observations}

\begin{figure}[h!]
\begin{center}
\includegraphics[width=16cm]{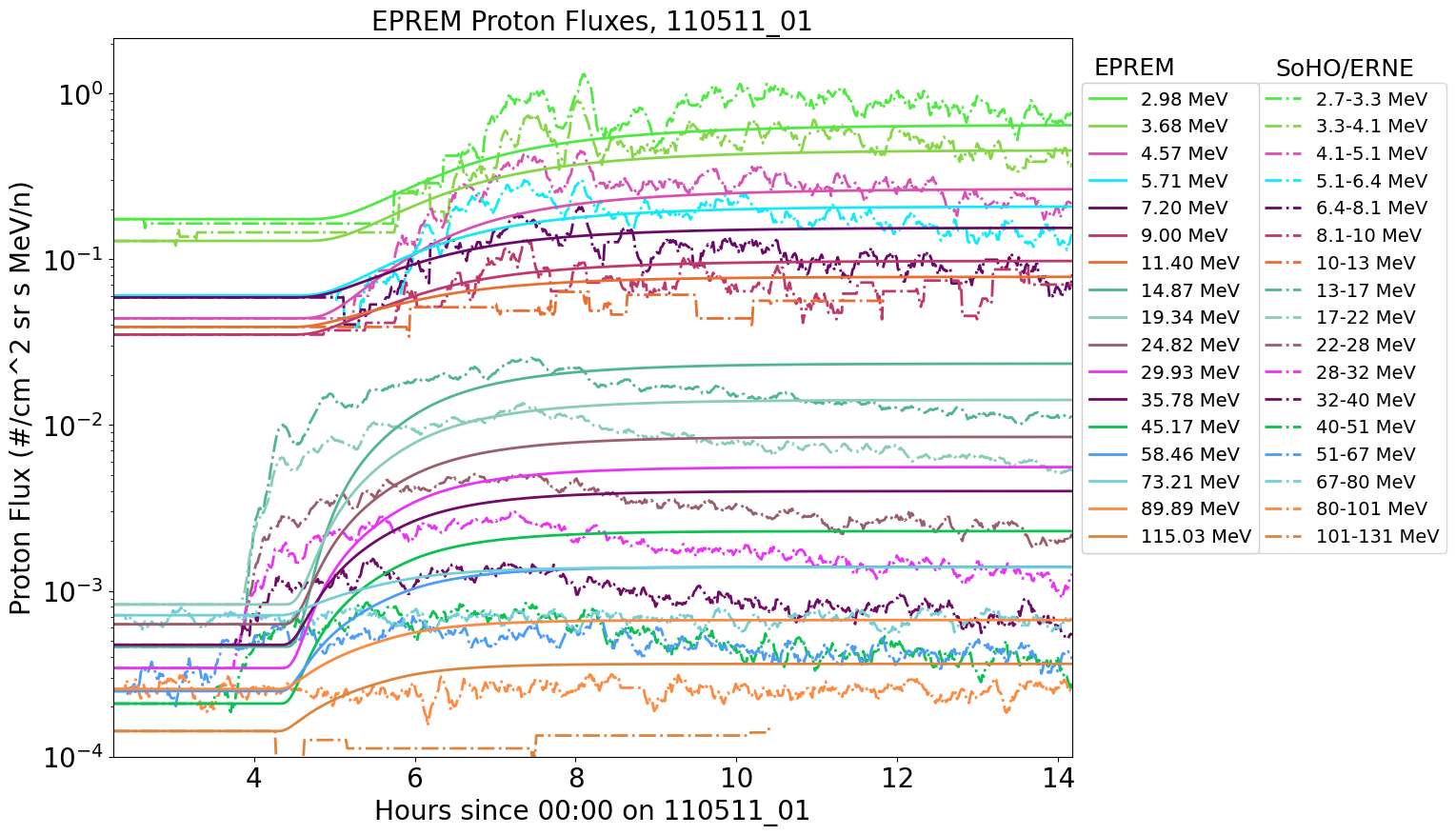}
\end{center}
\caption{The differential proton flux output of the EPREM model for the 110511 event near the L1 point. EPREM flux time series at the geometric mean energies of the ERNE channels are shown as colored solid lines. The ERNE observations are shown for comparison as dot-dashed lines colored the same as the corresponding EPREM energies. Time on the x-axis is in hours from the start of May 11, 2011.}
\label{fig_eprem_erne_fluxes_110511}
\end{figure}

Figure \ref{fig_eprem_erne_fluxes_110511} shows the output of the EPREM model for the  event. Proton differential fluxes as a function of time in hours are shown for the geometric mean energies of all ERNE channels between 3 and 115 MeV. The ERNE observations are also shown for direct comparison with dot-dashed lines, and corresponding colors for each energy channel. In order to improve the calculations of onset times and fluences, we add to the energetic particle model output the average pre-event flux from the ERNE observations. It is calculated by averaging the first five time steps of the ERNE fluxes in each energy channel.

\begin{figure}[h!]
\begin{center}
\includegraphics[width=16cm]{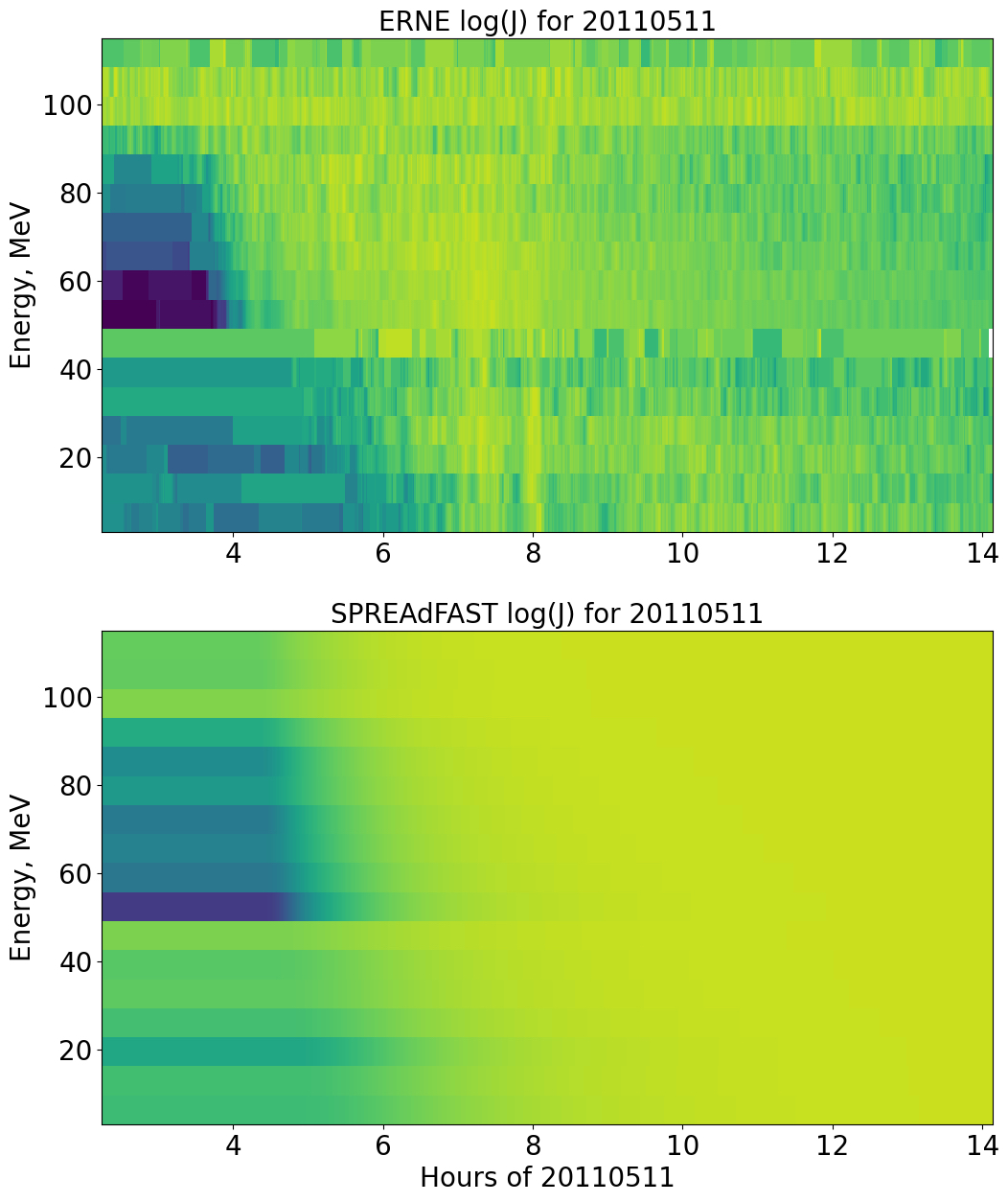}
\end{center}
\caption{Dynamic spectra of the differential proton fluxes of the ERNE observations (top) and the EPREM simulation (bottom) for the 110511 event. Time on the x-axis is in hours from the start of May 11, 2011.}
\label{fig_eprem_erne_dynamic_spectra_110511}
\end{figure}

Overall, for this event the model yields good agreement for the flux levels at most energies, compared with the observations. At the low-energy channels the agreement is best, including the onset of the particle event. Above 15~MeV, there is a discrepancy in the time profile, with the observed proton fluxes rising $\sim$1~hr before the simulation. This can also be seen in a more detailed comparison in Figure \ref{fig_eprem_erne_dynamic_spectra_110511}. It shows dynamic spectra of the ERNE and EPREM flux intensities. The fluxes in each energy channel have been normalized to the maximum ERNE flux values in it. The pre-event ERNE flux background is also included, for direct comparison to observations.
Such discrepancies will be addressed in future work, and likely corrected with improvements to the SPREAdFAST model chain. However, it is a testament to the strength of our method that such good overall agreement has been achieved with an initial set of input parameters, given the extremely complex nature of the problem and the many unknown parameters.

\begin{figure}[h!]
\begin{center}
\includegraphics[width=12cm]{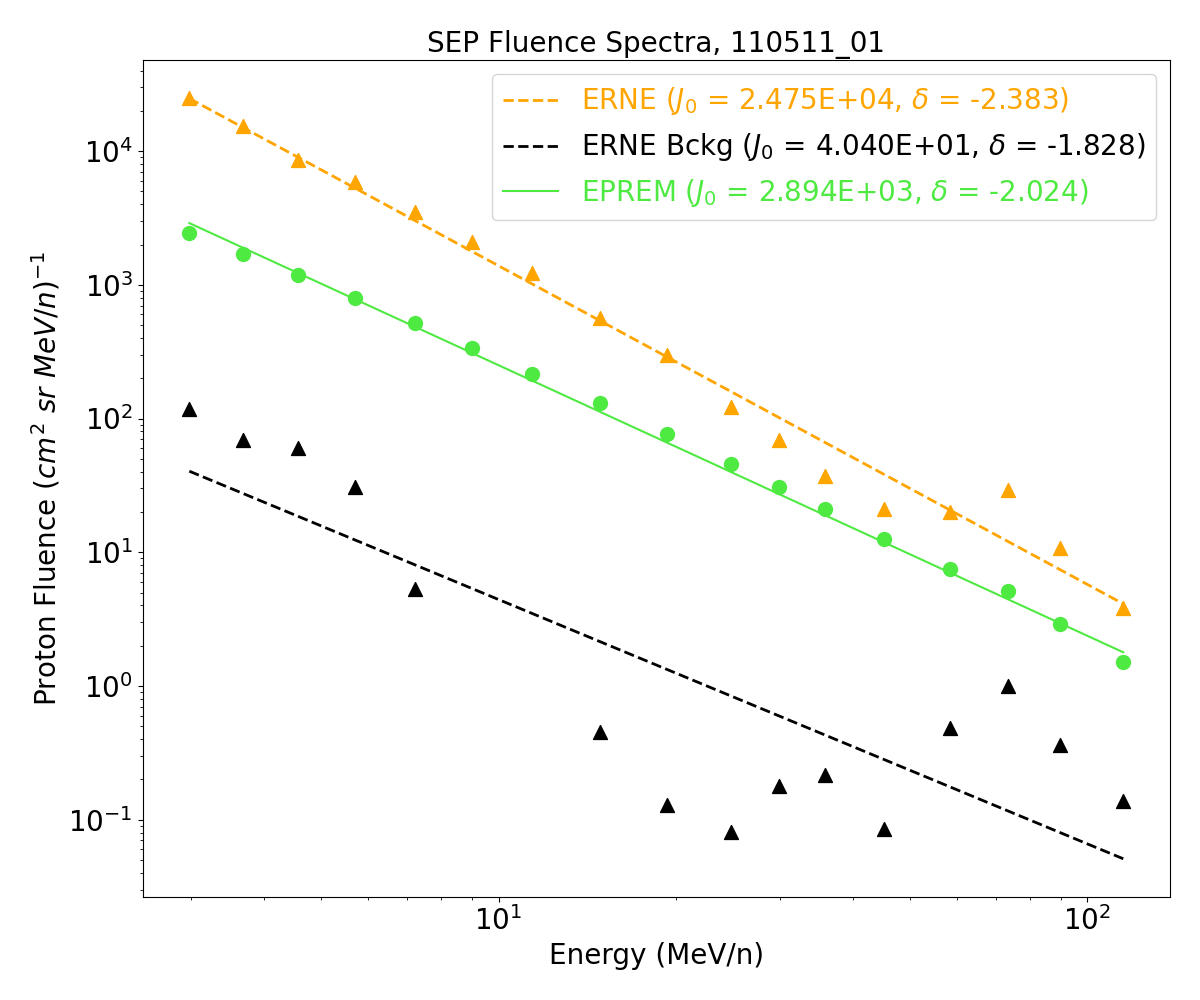}
\end{center}
\caption{A comparison between the modeled (green) and observed (yellow) proton fluences for the  event. The pre-event background ERNE spectrum is also shown with black symbols. Power law fit amplitudes and indices for all fluence spectra are shown as well.}
\label{fig_eprem_erne_fluences_110511}
\end{figure}

In Figure \ref{fig_eprem_erne_fluences_110511} we show a comparison between the event-integrated fluence spectra from the EPREM model and ERNE observations for the same event. The agreement is somewhat close, in terms of low-energy amplitude, and in terms of the fitted power law. In terms of radiation hardness, the EPREM spectrum is slightly flatter, which is desirable when making forecasts. In other words, in general it is better to forecast a harder spectrum than the observed, rather than a softer spectrum. This comparison between modeled and observed spectra is one of the metrics we discuss next as a way to evaluate the model performance over the full set of 62 events.

\begin{figure}[h!]
\begin{center}
\includegraphics[width=17cm]{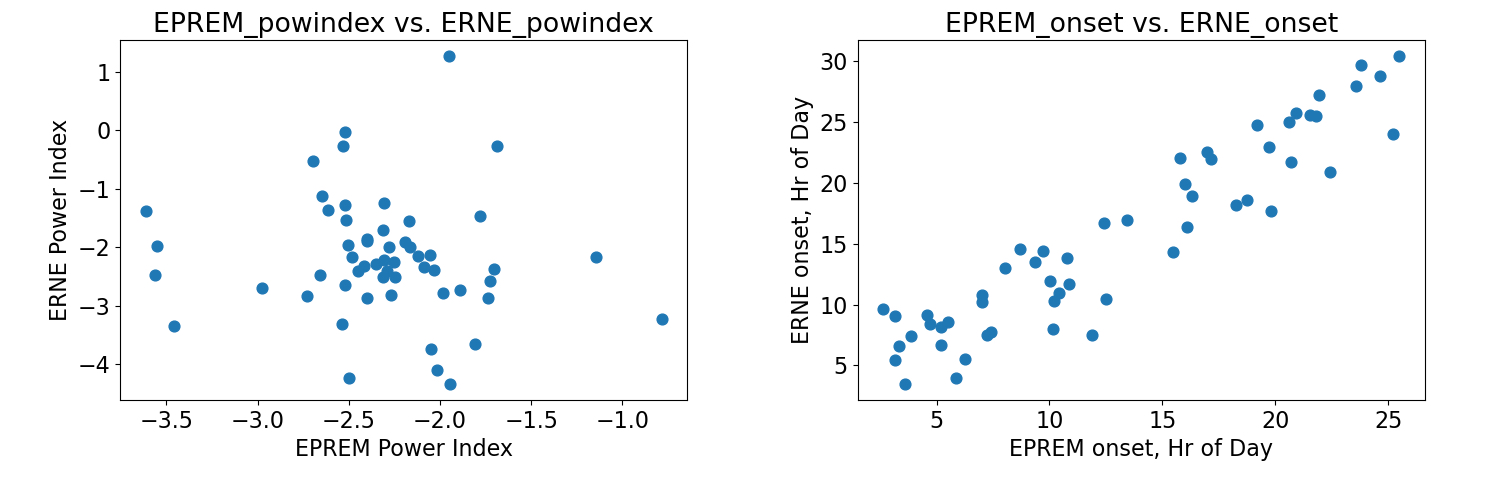}
\end{center}
\caption{Scatter plots of the relationships between the fitted power index (left) for all 62 events, as well as the onset times (right) for EPREM model and ERNE observations.}
\label{fig_eprem_erne_scatterplots}
\end{figure}

We have further explored the relations between the modeled and observed SEP parameters at 1 au for the full set of 62 studied events. Figure \ref{fig_eprem_erne_scatterplots} shows scatter plots of the relationships between the fitted power indices of the proton fluences from the EPREM model and the ERNE observations (left panel), and the onset hours for the two sets (right panel), defined as the time of 10\% increase above the average pre-event flux at the energy channel centered on 28~MeV. This energy is high enough that it is above the geometric mean of the ERNE energy range, but also low enough that flux enhancements are observed at this energy for most events. On the left panel, although the power law indices are centered around -2.5 for both EPREM and ERNE, there is considerable scatter, with a number of outliers, which we will investigate in future work. It is likely due to the input spectral shape and the scattering mean free path.

\begin{figure}[h!]
\begin{center}
\includegraphics[width=17cm]{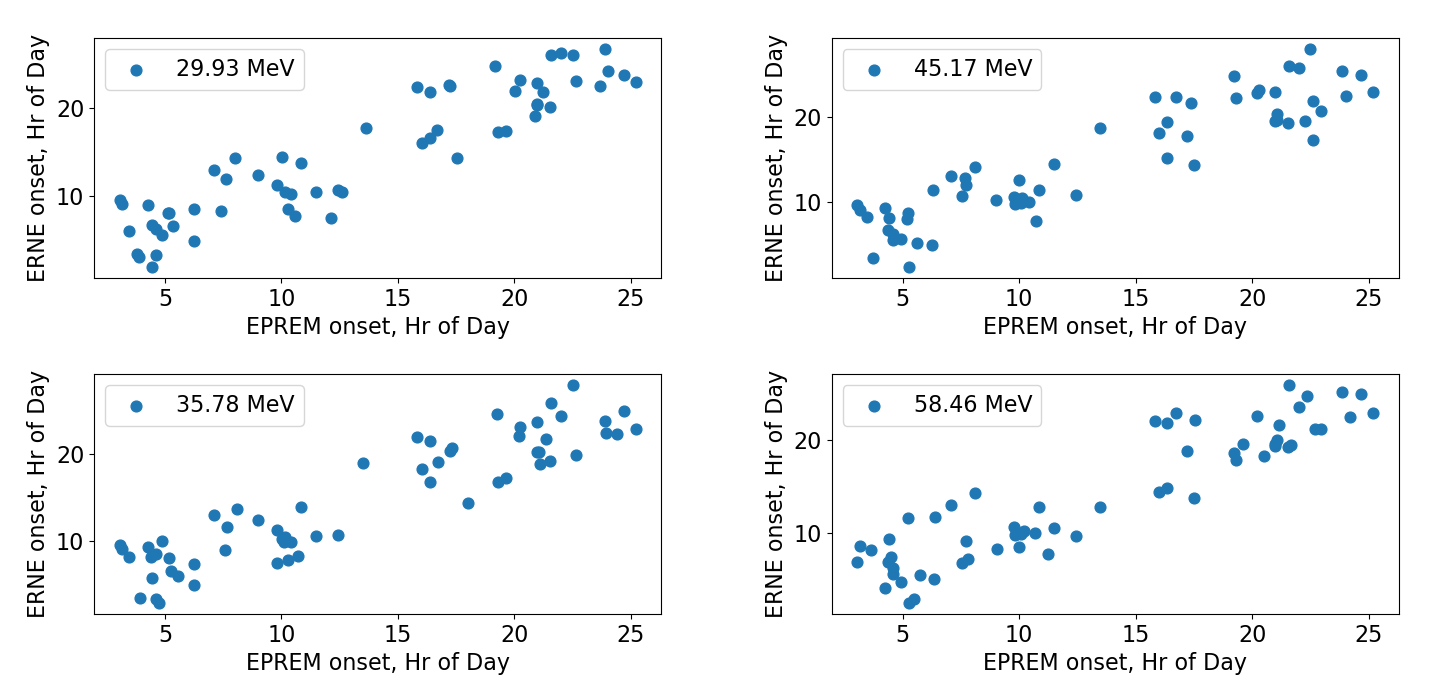}
\end{center}
\caption{Scatter plots of the onset times at four separate energies for EPREM model and ERNE observations, similar to the right plot in Fig. \ref{fig_onset_scatterplots}.}
\label{fig_onset_scatterplots}
\end{figure}

In the right panel, we find a quite strong linear relation between the onset times at 28~MeV, which gives us confidence in the modeling framework's performance. We further demonstrate the onset relationships for four higher energies in Figure \ref{fig_onset_scatterplots}. In future work, we will focus on narrowing the deviation of onset times from the observations by modifications in the input SEP spectra and the transport parameters. We note also that the EPREM model currently uses a nominal Parker spiral model for the interplanetary magnetic field; a realistic interplanetary model would likely change the length of the large-scale magnetic field lines along which the protons travel, due to meandering effects \citep{Laitinen:2018, Laitinen:2019}.

\begin{figure}[h!]
\begin{center}
\includegraphics[width=16cm]{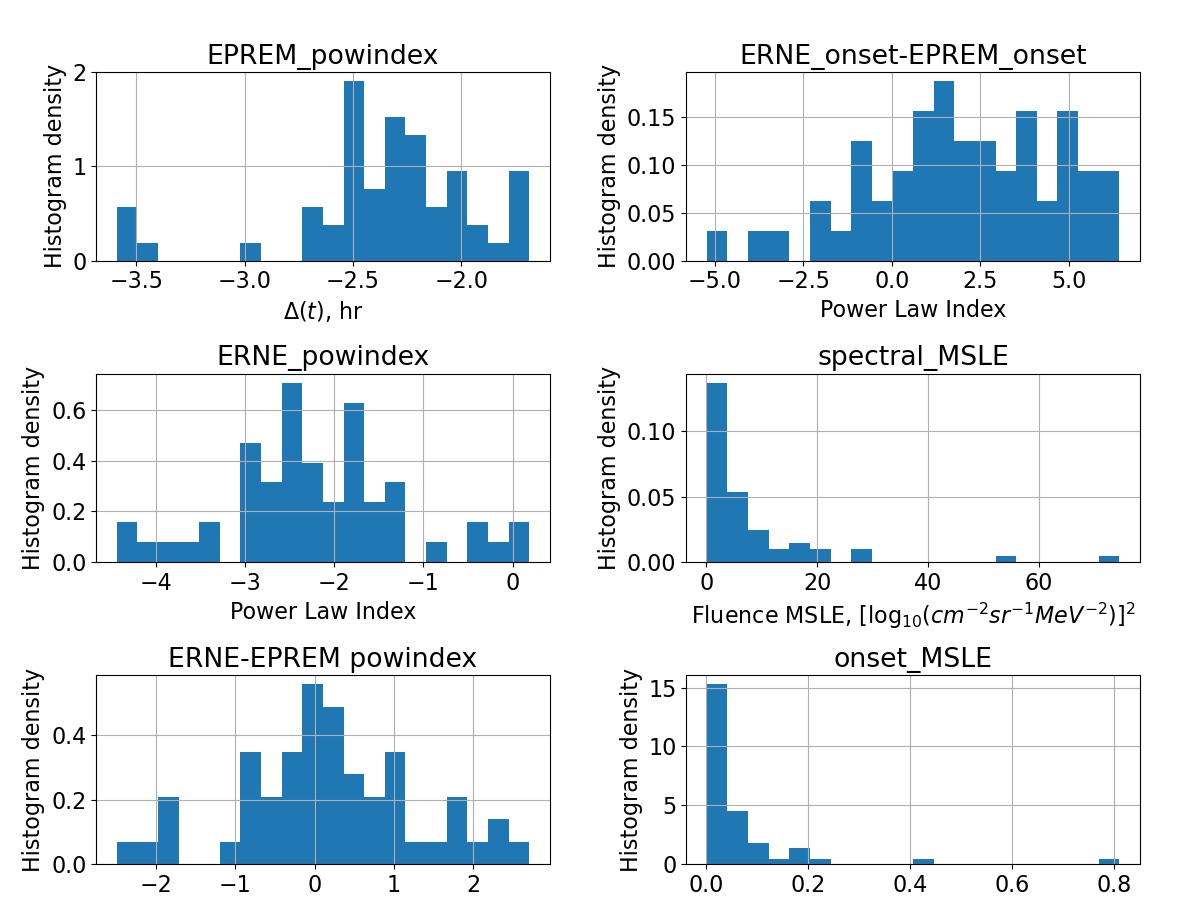}
\end{center}
\caption{A comparison between the modeled (yellow) and observed (blue) proton fluences for the  event. Power law fit amplitudes and indices for both fluence spectra are shown as well. The pre-event background ERNE fluence, together with the power law fit is also shown for reference.}
\label{fig_eprem_erne_histograms}
\end{figure}

We have also computed and show in Figure \ref{fig_eprem_erne_histograms}: histograms of the difference in hours between the modeled and observed flux onset at 28~MeV; the modeled (EPREM\_powindex) and observed (ERNE\_powindex) index of the power law fits to the resulting fluence spectra; and the Mean Squared Logarithmic Error (MSLE) for the fluence spectra and the event onsets. One of the most important parameters is the onset time - it is desired that the modeling framework predicts the proton fluxes arriving at 1 au before rather than after the actual onset, if it is to be used for forecasting. In this first study, we find that for about two-thirds of the events the modeled onset is before or close to the observed time (positive hour values in the top-left panel). The rest of the events will be further investigated in future work.

Comparing the EPREM\_powindex and ERNE\_powindex, we see an overall agreement of the peaks and extent of the distributions, although the simulations must be improved still. The last two panels show that the spectral and onset errors are sharply peaked near 0, meaning that for most events the resulting spectra are very similar, and the onsets are very close. This is a very positive result, demonstrating the capability of the framework. The individual flux and fluence comparisons are available at the SPREAdFAST catalog web page for all 62 events.

We find that in 2/3 of the events there is good agreement in the fluxes between model and observations. However, discrepancies were found in a number of events. This points to a likely existence of additional seed population beyond the quiet-time suprathermal spectra, possibly from flare-accelerated protons. We further find significant agreement with the observed proton flux profiles in the early stages for energies below 12 MeV. However, the model often produces higher fluxes than observations at the energies above $\sim30$~MeV, suggesting an energy dependence of the scattering mean free path, which must be investigated.

For most of the events, the ERNE fluxes continue rising for the full duration of the interplanetary simulation after onset, indicating continued injection of particles near the Sun, in line with the model setup. For some events, the flux increases are short-lived, resembling impulsive injections convolved with the transport effects. For other events still, there is a relatively quick increase and drop, followed by more prolonged increase. This may be due to flux dropout effects, and will be further investigated in future work.



\section{Conclusions}
\label{S4}

We have presented the first of its kind multi-event study of detailed physics-based Sun-to-1 au SEP simulations based on coronal diffusive shock acceleration and interplanetary propagation, of which we are aware. In this study, we investigated 62 separate eruptive events, which included an EUV CBF and sizable enhancements of 1 au proton fluxes. All but one had an associated flare. We applied the SPREAdFAST framework to model the evolution of the plasma immediately upstream of the coronal shock associated with the CBFs between 1 and 10~$R_{Sun}$, as well as the resulting acceleration and transport of protons between 1.05~$R_{Sun}$ and 1 au. The SPREAdFAST framework, intended for forecasting the early-stages of SEP events, has proved very useful in the study of such a large set of events.

The input spectra for the coronal proton acceleration were derived from quiet-time suprathermal spectra averaged over one to three days preceding each event; the resulting averages were scaled back to the Sun assuming a simple inverse square proportionality to heliospheric distance, as in \cite{Kozarev:2019}. We used an energy-independent mean free path for the coronal acceleration of the protons in the diffusive shock acceleration model of \cite{Kozarev:2016}. Similarly to previous work \citep{Kozarev:2015, Rouillard:2016, Kong:2017, Kong:2019}, we found that the conditions in the solar corona influence significantly the acceleration of protons. Specifically, the large gradients in the plasma parameters between neighboring streamers, quiet-Sun areas, and coronal holes have the effect of continuous, smooth changes in the $theta_{BN}$ angle along the shock wave surface, as well as in density and density enhancements. \cite{Kong:2019} found that the SEPs accelerated at a global shock wave expanding through a streamer in the low and middle corona will concentrate closer to the streamer apex, the higher the particle energy. However, the lower energy ($<10~MeV$) protons were distributed more or less uniformly along the shock surface in their 2D model. Thus, answering the question 'Where are SEPs produced?' in the early stages of CME-driven shock and compressive waves, depends strongly on the overlying coronal structure, and the particle energy.

The output from the DSA model was used as time-dependent input to the interplanetary transport EPREM model, which produced as input proton fluxes at 1 au. These were then compared with in situ observations by the SOHO/ERNE instrument. Overall, the results are very encouraging for the efficiency and accuracy of the SPREAdFAST model chain. However, we find that the fluxes at the highest energies show the most disagreement, due mostly to the slope of the increase and the onset times. Fig. \ref{fig_eprem_erne_histograms} shows an example of this for the May 11, 2011 event.

In future work, we will carry out more realistic modeling of the events in order to improve the match to observations. First, we will investigate how introducing time-dependent injection of the source spectra at the inner boundary of the EPREM simulation may change the L1 output, especially for those events in which the observed fluxes decay much faster than the modeled ones, or there are other discrepancies with observations. Although the focused transport is dominant, \citet{Dalla:2020} show that 3D transport effects in a realistic interplanetary magnetic fields are quite important. In future work we will implement perpendicular transport to account for such effects. We will also include location-dependent output, which does not guarantee constant connectivity between the source and observer. Finally, constraining the geometric shock models with existing (such as SOHO/LASCO) and new observations of the CME evolution in the middle corona will help reduce uncertainty in the results. Comparing near-Sun in situ observations of the quiet-time suprathermal populations by Parker Solar Probe and Solar Orbiter, and comparing them with the 1 au fluxes, will help us improve the input spectra estimation.

\section*{Conflict of Interest Statement}
The authors declare that the research was conducted in the absence of any commercial or financial relationships that could be construed as a potential conflict of interest.


\section*{Funding}
This work has been supported by ESA's the PECS Programme for Bulgaria, through Contract 4000126128 to the Institute of Astronomy and NAO, Bulgarian Academy of Sciences.


\section*{Supplemental Data}
Detailed results of the study for all events are available at the SPREAdFAST catalog: \href{http://spreadfast.astro.bas.bg/catalog/}{http://spreadfast.astro.bas.bg/catalog/}.




\begin{center}
\begin{longtable}{cccccc}
\caption{Events in this study}

\label{table:1} \\

\hline \multicolumn{1}{c}{\textbf{Event Date}} & \multicolumn{1}{c}{\textbf{Flare Start (UT)}} & \multicolumn{1}{c}{\textbf{Flare End (UT)}} &
\multicolumn{1}{c}{\textbf{Flare Class}} & \multicolumn{1}{c}{\textbf{Source X ($"$)}} & \multicolumn{1}{c}{\textbf{Source Y ($"$)}} \\
\hline 
\endfirsthead
\multicolumn{6}{c}%
{{\bfseries \tablename\ \thetable{} -- continued from previous page}} \\
\hline \multicolumn{1}{c}{\textbf{Event Date}} & \multicolumn{1}{c}{\textbf{Flare Start (UT)}} & \multicolumn{1}{c}{\textbf{Flare End (UT)}} & 
\multicolumn{1}{c}{\textbf{Flare Class}} & \multicolumn{1}{c}{\textbf{Source X ($"$)}} & \multicolumn{1}{c}{\textbf{Source Y ($"$)}}  \\
\hline 
\endhead
\hline \multicolumn{6}{c}{{Continued on next page}} \\ 
\hline
\endfoot
\hline \hline
\endlastfoot
2010/06/12 & 00:53 & 01:19 & M2.0 & 633 & 390 \\
2010/08/07 & 18:05 & 18:50 & M1.0 & -455 & 59 \\
2010/08/14 & 09:30 & 10:08 & C4.4 & 697 & -26 \\
2010/08/18 & 05:12 & 05:50 & C4.5 & 987 & 342 \\
2010/12/31 & 04:15 & 05:01 & C1.3 & 799 & 246 \\
2011/01/28 & 00:45 & 01:59 & M1.3 & 949 & 218 \\
2011/03/07 & 19:34 & 20:30 & M3.7 & 614 & 553 \\
2011/05/11 & 02:20 & 02:35 & B8.1 & 785 & 399 \\
2011/06/07 & 06:15 & 06:40 & M2.5 & 611 & -355 \\
2011/08/03 & 13:20 & 14:00 & M6.0 & 454 & 195 \\
2011/08/04 & 03:43 & 04:20 & M9.3 & 546 & 200 \\
2011/08/08 & 17:45 & 18:43 & M3.5 & 812 & 215 \\
2011/08/09 & 07:50 & 08:30 & X6.9 & 866 & 229 \\
2011/09/04 & 04:32 & 05:07 & C9.0 & 893 & -323 \\
2012/01/27 & 17:56 & 18:38 & X1.7 & 857 & 454 \\
2012/03/04 & 10:25 & 11:10 & M2.0 & -833 & 332 \\
2012/03/07 & 00:00 & 00:40 & X1.3 & -475 & 397 \\
2012/03/13 & 17:03 & 17:44 & M7.9 & 804 & 352 \\
2012/04/05 & 20:50 & 21:30 & C1.5 & 487 & 364 \\
2012/04/09 & 11:54 & 12:40 & M3.9 & 812 & 382 \\
2012/07/19 & 04:45 & 05:30 & M7.7 & 917 & -217 \\
2012/07/23 & 02:09 & 02:48 & U-FL & 912 & -243 \\
2012/08/31 & 19:15 & 21:13 & C8.4 & -641 & -444 \\
2012/11/08 & 02:05 & 02:50 & M1.7 & -931 & 226 \\
2013/04/11 & 06:53 & 07:25 & M6.5 & -245 & 260 \\
2013/04/21 & 06:35 & 07:35 & U-FL & 937 & 181 \\
2013/05/13 & 15:44 & 16:20 & X2.8 & -927 & 186 \\
2013/05/15 & 01:06 & 01:50 & X1.2 & -852 & 199 \\
2013/05/22 & 12:33 & 13:20 & M5.0 & 875 & 238 \\
2013/06/21 & 02:31 & 03:21 & M2.9 & -869 & -268 \\
2013/09/29 & 21:18 & 22:25 & C1.2 & 420 & 166 \\
2013/10/11 & 07:01 & 07:41 & M1.5 & -897 & 342 \\
2013/10/25 & 07:53 & 08:29 & X1.7 & -914 & -158 \\
2013/12/07 & 07:17 & 07:36 & M1.2 & 690 & -270 \\
2013/12/12 & 03:03 & 03:33 & B2.2 & 750 & -450 \\
2013/12/28 & 17:10 & 18:00 & C9.3 & 942 & -252 \\
2014/01/06 & 07:38 & 08:17 & U-FL & 942 & -252 \\
2014/01/07 & 17:58 & 18:37 & X1.2 & 198 & -159 \\
2014/02/20 & 07:23 & 08:07 & M3.0 & 897 & -216 \\
2014/02/25 & 00:38 & 01:14 & X4.9 & -939 & -184 \\
2014/03/28 & 19:13 & 20:00 & M2.6 & 339 & 285 \\
2014/04/02 & 13:03 & 13:47 & M6.5 & -745 & 294 \\
2014/04/25 & 00:15 & 00:49 & X1.3 & 929 & -214 \\
2014/05/09 & 02:15 & 02:59 & U-FL & 936 & -165 \\
2014/07/08 & 16:06 & 16:51 & M6.5 & -767 & 163 \\
2014/08/25 & 14:11 & 14:51 & M2.0 & 581 & 40 \\
2014/08/28 & 15:06 & 15:47 & C1.9 & 921 & 107 \\
2014/09/10 & 17:20 & 18:15 & X1.6 & -32 & 113 \\
2014/10/02 & 18:48 & 19:30 & M7.3 & 908 & -293 \\
2014/12/05 & 05:42 & 06:21 & C2.1 & 872 & -366 \\
2014/12/13 & 14:01 & 14:41 & U-FL & 915 & -333 \\
2015/05/12 & 02:18 & 02:49 & C2.6 & 960 & -192 \\
2015/09/20 & 17:28 & 18:11 & M2.1 & 660 & -429 \\
2015/10/29 & 02:13 & 02:52 & U-FL & 951 & -167 \\
2015/11/09 & 12:51 & 13:27 & M3.9 & -626 & -229 \\
2016/01/01 & 22:42 & 23:59 & M2.4 & 903 & -325 \\
2016/03/16 & 06:33 & 07:14 & C2.2 & 943 & 199 \\
2016/04/18 & 00:17 & 01:08 & M6.7 & 826 & 239 \\
2017/04/01 & 21:31 & 22:19 & M4.4 & 761 & 308 \\
2017/04/03 & 14:13 & 15:09 & M5.8 & 894 & 302 \\
2017/04/18 & 19:15 & 20:35 & C5.5 & -920 & 241 \\
2017/09/02 & 15:16 & 16:59 & C7.7 & 948 & 66\\
\hline
\end{longtable}
\end{center}
\end{document}